# Complete spin phase diagram of the fractional quantum Hall liquid

H. M. Yoo[1], K. W. Baldwin[2], K. West[2], L. Pfeiffer[2], R. C. Ashoori[1]

[1]Department of Physics, Massachusetts Institute of Technology, Cambridge, MA 02139, USA

[2]Department of Electrical Engineering, Princeton University, Princeton, NJ 08544, USA

Study of the ground-state electronic spin-polarization can permit discovery and identification of novel correlated phases in the quantum Hall (QH) system. It can thus determine the potential usefulness of QH states for quantum computing. However, prior measurements involving optical and NMR techniques may perturb the system away from delicate ground-states. Here, we present spin-resolved pulsed tunneling (SRPT) that precisely determines the complete phase diagram of the ground-state spin-polarization as a function of magnetic field ($B$) and filling factor ($\nu$). We observe fully-polarized $\nu = 5/2$ and 8/3 states at very small $B$, suggesting strong deviation from the weakly-interacting composite-fermion model that largely successfully describes our phase diagram for the lowest Landau level. The results establish SRPT as a powerful technique for investigating correlated electron phenomena.

The fractional quantum Hall effect (FQHE) occurs in 2D electronic systems (2DES) in the presence of a perpendicular magnetic field. In the FQHE, electrons exist in a many-body wavefunction that minimizes their mutual Coulomb interactions and, remarkably, leads to excitations with a fraction of the electron charge (*1, 2*). The spin-degree of freedom of the electrons adds to the complexity to the FQHE and can also generate fundamentally new phases. For instance, spin-singlet and topological spin phases emerge from the addition of a spin degree of freedom as the strength of the Coulomb interaction $e^2/\epsilon l_B$ becomes comparable to the single-particle Zeeman energy $E_z$.

Experimenters have made substantial efforts to probe the spin-polarization of electrons in the FQHE system. NMR, optical, and transport measurements, have resulted in evidence of numerous spin phase transitions in the lowest ($N = 0$) Landau Level (LL) (*3–7*). However, the NMR and optical methods used in previous measurements of spin-polarization may perturb the system away from delicate ground-states (*8*). Transport measurements do not directly measure the spin-polarization and depend on modelling for interpretation. Finally, all previous methods have significant limitations in the LL filling factors that they can probe (*8*). Here, we introduce spin-resolved pulsed tunneling (SRPT). SRPT offers a unique approach to probe the spin properties of a quantum Hall system. By using electrical pulses that drive tunneling over very short time intervals, it avoids undesired perturbations such as heating. Pulsed tunneling also provides a capability critical to the functioning of SRPT: it permits probing of electron tunneling into and out of insulating ferromagnetic states (*9*). The high-throughput and dramatically improved signal-to-noise ratio of SRPT measurements allow precise determination of the ground-state spin polarization over a wide range of filling factors $\nu$ and magnetic fields as low as $B_\perp \sim 1$ T. In this report, we describe SRPT and results from it that show a breakdown of the conventional theory of the FQHE in the composite fermion (CF) picture in the $N = 1$ LL as well as the delicate skyrmionic states in the FQHE regime in the $N = 0$ LL.

In order to perform SRPT measurements, we create a magnetic tunnel junction consisting of a "probe" 2DES fixed in a ferromagnetic state at LL filling factor $\nu = 1$ and another layer (the "2DES under study") with tunable $\nu$ (Fig. 1C). For a 2DES at filling factor $\nu = 1$, strong exchange interactions drive a large splitting between spin-up and spin-down states. This system, a "QH ferromagnet", behaves as a fully polarized ferromagnet, with all spin-up states in the Landau level filled and all spin-down states empty (*10, 11*). We can thus drive spin-resolved tunneling between the layers. For instance, when the layer under study is at $\nu \leq 2$, only spin-down electrons can tunnel into the QH ferromagnet because there are no available states for spin-up electrons in the $N = 0$ LL in the QH ferromagnet. Figure 1C shows tunneling current spectra measured in the 2D-QH ferromagnet tunnel junction. In contrast to the spin-degenerate spectra in Fig. 1B, the spectra show significant changes, such as a sharp suppression of tunneling current near odd fillings where the 2D-system under study is spin-polarized. We also observe notable differences between the spin-degenerate (2D-2D) and spin-resolved (2D-QH ferromagnet) tunneling at fractional fillings, in which electrons are condensed into the spin-polarized or -unpolarized ground states (Fig. 1D and E).

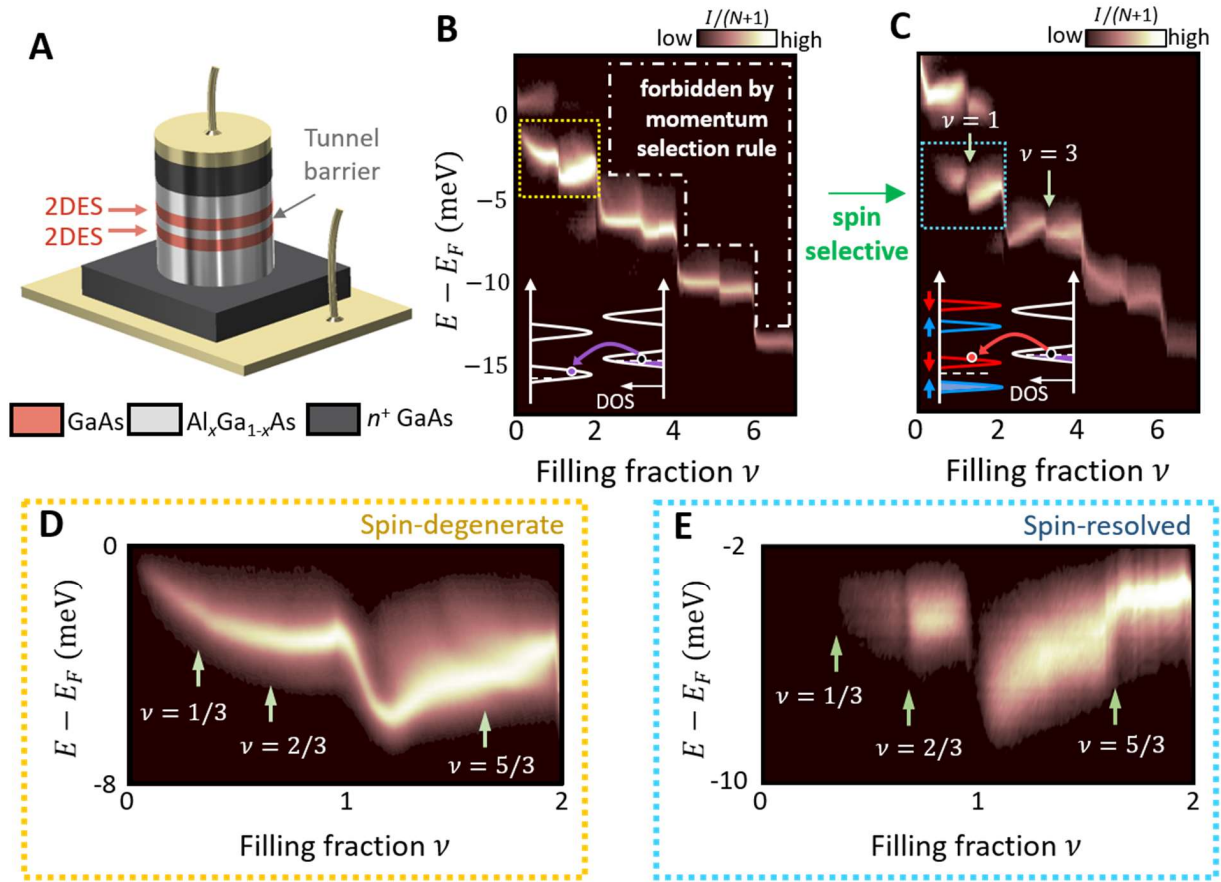

**Fig. 1. Tunneling in the quantum Hall effect regime.** (**A**) Schematic view of a 2D-2D tunnel device (*8*). (**B and C**) Tunneling spectra of the (B) spin-degenerate 2D-2D and (C) 2D-QH

ferromagnet tunnel junctions measured at a fixed $B_\perp = 2$ T. Insets show the schematic energy diagram of the tunnel processes. Vertical axis is the tunneling energy $E$ referenced to the Fermi energy $E_F$. Horizontal axis is $v$ proportional to the electron density ($n$) of the 2D system under study. The color scale is the current $I$ scaled by the topmost occupied LL index ($N + 1$). **(B)** The spin-degenerate 2D-2D tunnel junction consists of an empty ($v = 0$) layer and a 2D layer with tunable $n$ and $v$ (the 2DES under study). The spin-degenerate tunneling spectra show a single equidistant staircase pattern (see Fig. S6 and S7 for detailed description). **(C)** An additional spin-selection rule applies to a magnetic tunnel junction consisting of $v = 1$ QH ferromagnet and the 2DES under study. **(D and E)** The selected region of the 2D-2D and 2D-QH ferromagnet tunneling spectra at a higher $B_\perp = 5.2$ T. A sharp increase and decrease of $I$ occurs at $v = 1/3$, 2/3, and 5/3 in (E), whereas no such feature exists in (D).

As only spin-down electrons from the 2DES under study can tunnel into $v = 1$ ferromagnetic probe layer, the integral of tunneling current is proportional to the spin-down electron density ($n_\downarrow$) occupied in the 2D system under study (*8*). Therefore, for a given filling factor $v$ of the 2DES under study, we can directly measure the bare spin polarization $P = 1 - \frac{2}{v}\left(\frac{n_\downarrow}{n_D}\right)$, where $n_D$ is LL degeneracy. We then deduce the normalized spin polarization $P^*$ that takes account of particle-hole duality in a LL.

$$P^* = \begin{cases} P & \text{for } v \leq 1 \\ \dfrac{v}{2-v} \times P & \text{for } 1 < v \leq 2 \end{cases} \qquad (1)$$

$P^*$ is the polarization relative to the maximum possible polarization. For the case of exact particle-hole symmetry between $v$ and 2 - $v$, Equation 1 yields a symmetric $P^*$ around $v = 1$.

Figure 2A shows $P^*$ measured from the integrated tunneling current at a fixed $B_\perp = 5.2$ T, with nearly full spin polarization at $v = 1$, as expected. We observe a sharp decrease of $P^*$ as the filling factor deviates from $v = 1$, consistent with the integer quantum Hall (IQH) skyrmion model (*13*), where the electron spins are flipped by adding or removing charges near $v = 1$. We also observe sharp changes of $P^*$ at fractional fillings, consistent with the CF model (*2*). In the CF model, the strongly interacting electrons at fractional fillings behave as nearly free quasiparticles formed from electrons with an even-number (2*p*) of magnetic flux quanta attached (see caption of Fig. 2B). Effective IQH states for the CFs develop at fractional $v$ that correspond to integer fillings of the CF LLs ($\Lambda$-levels). Within the $\Lambda$-levels, the spin-splitting increases with $E_z$ (grows linearly with $B_\perp$). Full spin-polarization occurs when $E_z$ grows larger than the splitting between adjacent $\Lambda$-levels, $\hbar\omega_c^*$, that scales with the repulsive Coulomb energy ($e^2/\epsilon l_B$). Therefore, $P^*$ increases with the ratio of $E_z$ to $e^2/\epsilon l_B$ that grows as $\sqrt{B_\perp}$. At a fixed $B_\perp$, $P^*$ varies with $v = v^*/(2pv^* - 1)$, where $v^*$ denotes the filling factor of the spin-split $\Lambda$-levels. The $v = 1/3$ state, for instance, corresponds to full-filling of the lowest spin-split $\Lambda$-level ($v^* = 1$) and is expected to be fully spin-polarized. On the other hand, the $v = 2/5$ state, corresponding to $v^* = 2$, is partially depolarized at small $E_z/(e^2/\epsilon l_B)$ (see Fig. 2C).

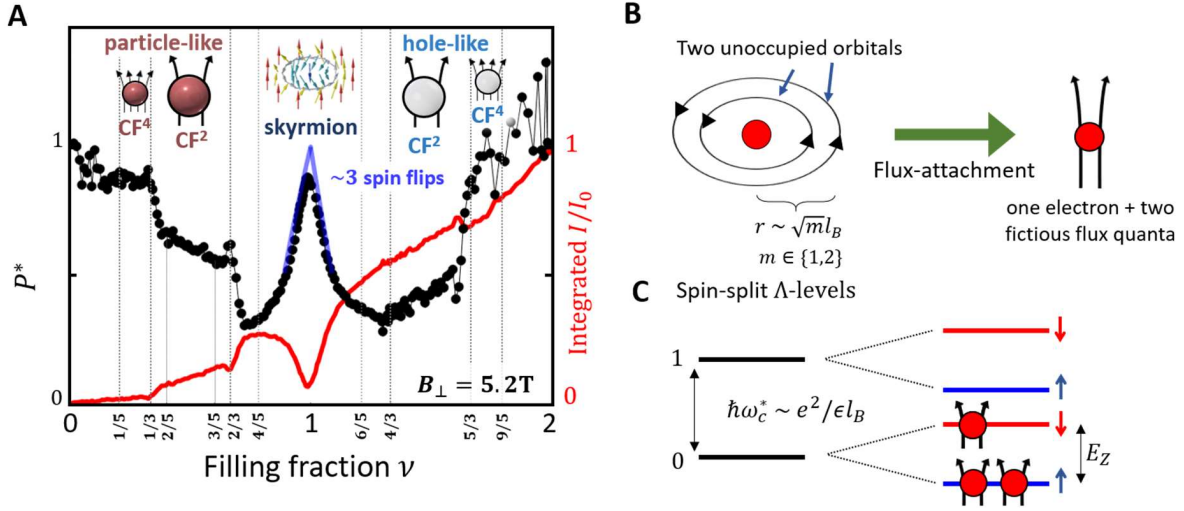

**Fig. 2. (A)** The integrated tunneling current (red) and the normalized spin polarization $P^*$ (black). Sharp changes of $P^*$ occur at integer and fractional fillings, where spin transitions between polarized and unpolarized states take place. **(B)** Cartoon picture of the flux attachment. Each electron orbit encloses one additional magnetic flux quantum $\phi_0$. At a fractional filling $\nu = 1/3$, the next two orbits around the central orbit are depleted due to strong repulsive interactions (*12*). The electron that carries two unfilled orbitals can be treated as a composite fermion (CF) consisting of one electron and the two fictitious flux quanta. **(C)** Schematic energy level diagram of the CF phase. The excited-state is separated from the ground-state by an energy gap that is proportional to $e^2/\epsilon l_B$ that produces the orbital configuration in (B). For a sufficiently large $E_z$, spin-splitting within the $\Lambda$-levels occurs.

Figure 3A shows the complete $P^*$ phase diagram, where changes in $B_\perp$ and $\nu$ drive phase transitions between the spin polarized, unpolarized, and partly polarized states. We observe that the composite-fermion model (*16*) of Fig. 3B qualitatively fits the major trends of the data. At large $B_\perp$, fully polarized states emerge because $E_z > \hbar\omega_c^*$. At small $B_\perp$, $P^*$ changes rapidly with $\nu$ because $\nu$ tunes both the strength of the effective magnetic field ($B_\perp^* = B - (2p)n\phi_0$) and the occupation difference between the spin-up and spin-down $\Lambda$-levels. Adding an additional magnetic field parallel to the plane of the 2D system also supports this trend of increasing polarization due to enhanced $E_z$ (Fig. S8).

Important details of the phase diagram, however, conflict with the free CF model. The discrepancy stems from the interactions between CFs in a partially filled $\Lambda$-level (*17*), analogous to a partially filled LL where the electron-electron interactions produce new collective states. For instance, the $P^*$ phase diagram shows the spin-unpolarized $\nu = 4/5$ and 6/5 states over a broad range of $B_\perp$ and $B_\parallel$. The spin-unpolarized FQH states at $\nu = 4/5$ and 6/5 (*14, 15*) cannot be explained by the particle-hole conjugates of the conventional FQH states at $\nu = 1/5$ and 9/5, respectively (see Fig. S9 and S10 for a detailed explanation). Furthermore, the $P^*$ phase diagram shows a sharp peak at $\nu = 1/3$ followed by rapid depolarization on both sides of $\nu = 1/3$ below 5 T. These features resemble the spin-depolarization near $\nu = 1$. In the CF model, $\nu = 1/3$ corresponds to $\nu^* = 1$, and a FQH skyrmion is expected as a quasiparticle is added or removed from $\nu = 1/3$ (18). Prior experiments (19, 20) measured an energy gap at $\nu = 1/3$, displaying features apparently indicative of the FQH skyrmion but that later theoretical work (21) suggests may result from disorder.

Our phase diagram provides two features in agreement with expectations for interacting CFs forming the FQH skyrmion. Firstly, unlike the IQH skyrmion near $\nu = 1$, the FQH skyrmion only appears at $B_\perp \lesssim 5$T. This agrees with the expectation that Coulomb driven exchange interactions of the CFs are substantially smaller than the inter-electron interaction ($e^2/\epsilon l_B$) at a fixed $B_\perp$ (18). Secondly, the FQH skyrmions display an asymmetry for the skyrmion ($\nu^* > 1$) and anti-skyrmion ($\nu^* < 1$) pair (22). The IQH skyrmion and anti-skyrmion pair are nearly symmetric because $e^2/\epsilon l_B$ is equal at $\nu > 1$ and $\nu < 1$ (23). However, the inter-CF interactions differ at $\nu^* > 1$ and $\nu^* < 1$ because, in the CF language, $B_\perp^*$ is larger at $\nu^* < 1$. The $P^*$ phase diagram in Fig. 3D corroborates such an asymmetric behavior, where the anti-skyrmion at $\nu^* < 1$ survives at a greater $B_\perp$.

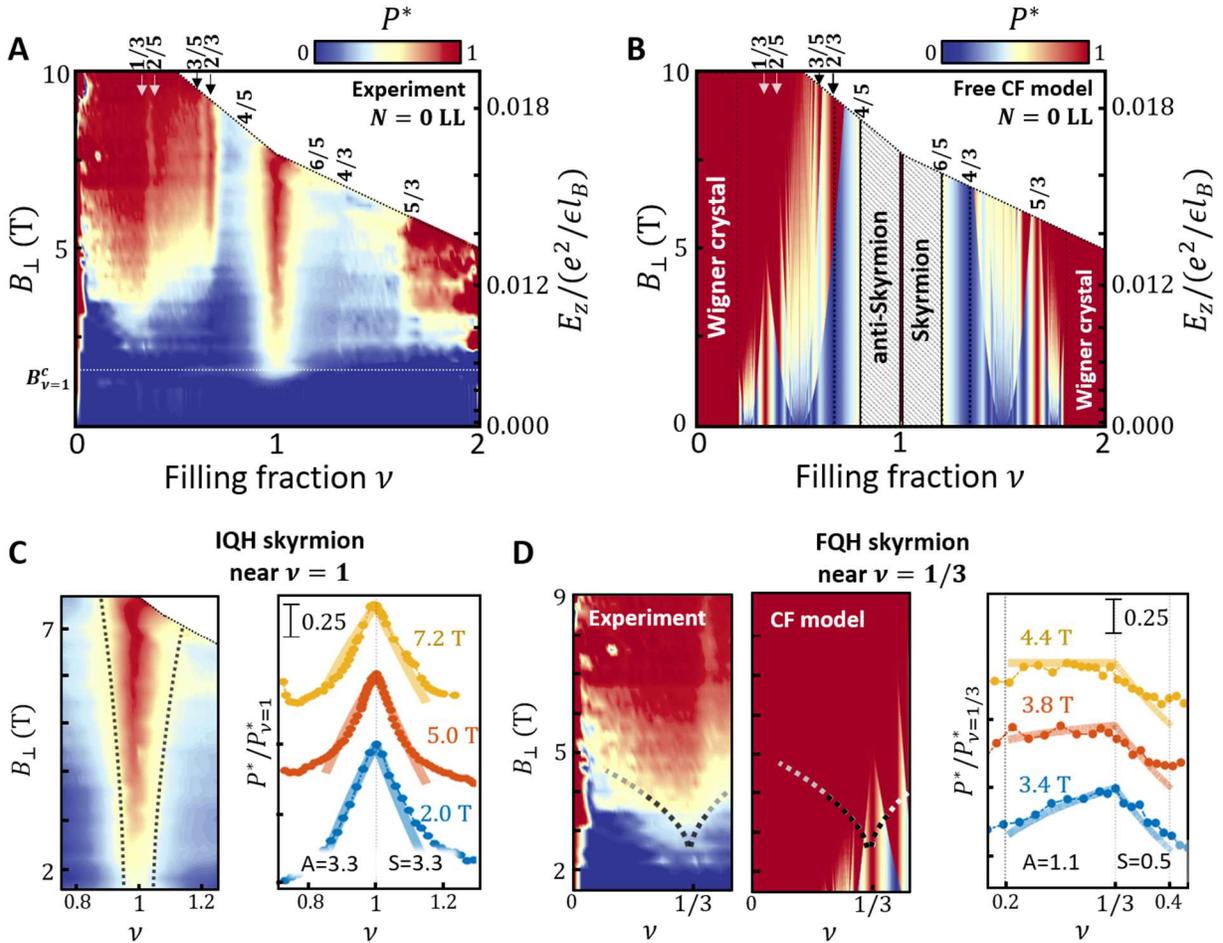

**Fig. 3. $B - \nu$ phase diagram of ground-state spin polarization in the $N = 0$ Landau level. (A)** The $P^*$ phase diagram measured as functions of $B_\perp$ and $\nu$. **(B)** Theoretical $P^*$ calculated from the free CF model (*16*). Effective masses $m_{CF}^* = 0.66 m_e \sqrt{B_\perp}$ at $v \le 1$ and $0.45 m_e \sqrt{B_\perp}$ at $v \ge 1$ are assumed in the model (see Fig. 4C for the origin of these CF mass values). Competing phases such as skyrmion and Wigner crystal phases are expected near integer values of $v$. Unusual interacting two-flux quantum CFs (*14, 15*) are also assumed at $2/3 < v \le 4/5$ and $6/5 \le v < 4/3$ (see Fig. S9 and S10). **(C and D)** Left panels show the $P^*$ phase diagrams in the vicinity of $v = 1$ and $v = 1/3$. Moving slightly away from $v = 1$ and $v = 1/3$ produces sharp spin depolarization. Dashed lines are guides to the eye. Right panels show line-cuts at several $B_\perp$. The shaded colored lines are fits based on the finite-size skyrmion model (*13*). $A$ and $S$ are the number of reversed spins in the anti-skyrmion and the skyrmion, respectively. Fully polarized $v = 1$ and $v = 1/3$ states are assumed in the model, an assumption that breaks down at a small $B_\perp$ due to sample inhomogeneity that depolarizes the $v = 1$ and $v = 1/3$ states.

We extend the $P^*$ measurement to the $N = 1$ LL. As the momentum selection rule forbids tunneling between the $N = 1$ LL in the 2DES under study and the $N = 0$ LL in the QH ferromagnet, we tilt the sample by 20° to allow tunneling between the two systems (*8*). We observed no qualitative changes of tunneling *I-V* characteristics at a smaller tilt angle below 20° (Fig. S11).

Figure 4A reveals the $P^*$ phase diagram measured in the $N = 1$ LL. In contrast with the $N = 0$ LL, the phase diagram appears completely different from that expected from the free CF model (Fig. 4B). In the free CF model, spin transitions from unpolarized states to fully spin-polarized states occur with increasing $B_\perp$ when $E_z$ grows larger than $\hbar \omega_c^*$. A previous transport experiment (*24*) demonstrated evidence of such spin-transitions at $v = 8/3$ leading to the conclusion that $v = 8/3$ is analogous to the conventional Laughlin-type $v = 2/3$ state (*24, 25*). In strong contrast, we observe no spin transitions, not only at the exact fractional fillings such as $v = 8/3$, but also at all other fillings between $v = 5/2$ and $v = 8/3$, where the free CF model predicts spin depolarization that should persist to larger $B_\perp$. In addition to the absence of the spin transitions, we observe anomalous spin depolarization over the range $11/5 \le v \le 7/3$. This behavior, again, cannot be explained by the CF model that predicts full spin polarization in this range.

To characterize the nature of the exotic behavior, we compare the CF effective masses $m_{CF}^*$ in $N = 0$ and $N = 1$ LLs. In the CF model, a half-filled LL is understood as a compressible (metallic) Fermi sea because the CFs experience zero $B_\perp^*$ (*26*). Pauli paramagnetic behavior of the Fermi sea is therefore predicted upon varying $E_z$ relative to the CF Fermi energy $E_F^* = (n_D \pi \hbar^2)/m_{CF}^*$, where $m_{CF}^*$ grows as $\sqrt{B_\perp}$ (*27*). We can thus determine $m_{CF}^*$ from the ratio of $E_z / E_F^*$. Figure 4C shows the $B_\perp$ dependence of $P^*$ at $v = 1/2$, 3/2, and 5/2. In the $N = 0$ LL ($v = 1/2$ and 3/2), $P^*$ is consistent with the Pauli paramagnetism model. In particular, the $m_{CF}^*$ extracted from the $B_\perp$ fits is in excellent agreement with the free CF model mass (*16*). On the other hand, $P^*$ at $v = 5/2$ is fully-polarized even at a very small $B_\perp$, indicating that $m_{CF}^*$ is dramatically enhanced in the $N = 1$ LL. The absence of spin transitions near $v = 8/3$ can be also understood in terms of the extremely heavy $m^*_{CF}$ that significantly reduces $\hbar\omega^*_c$.

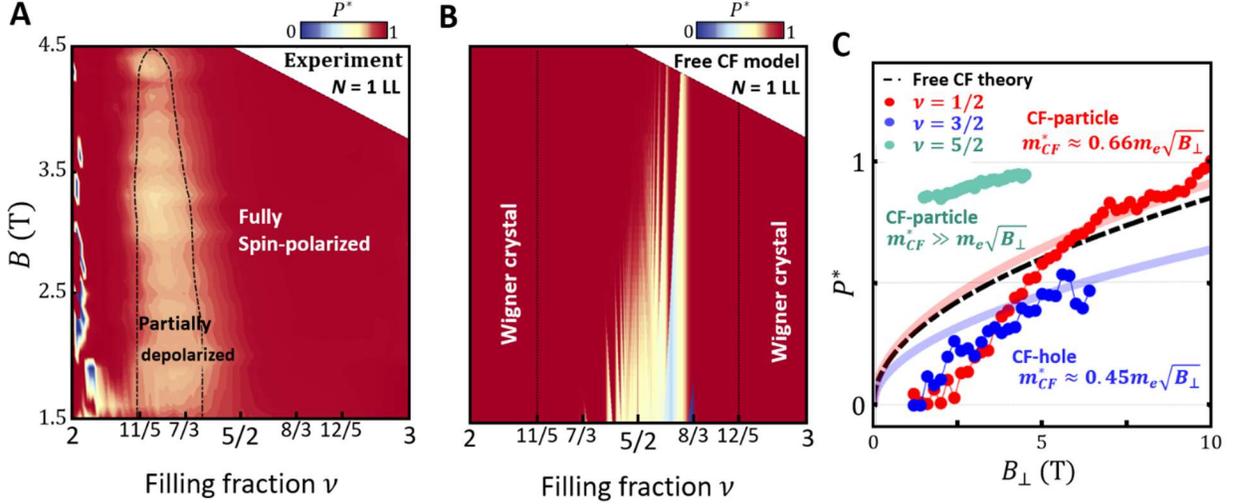

**Fig. 4. The ground-state spin polarization in the first excited $N = 1$ Landau level. (A)** The $P^*$ phase diagram measured in the $N = 1$ LL. The region enclosed in dashed line delineates partially depolarized spin states with $P^* < 0.8$. **(B)** The $P^*$ phase diagram of the free CF model using an effective mass $m^*_{CF} = 1.06 m_e \sqrt{B_\perp}$ (with $B_\perp$ in Tesla) at $v = 5/2$ obtained from the resistively-detected NMR study (*28*). **(C)** Comparison of $P^*$ and $m^*_{CF}$ in the half-filled $N = 0$ and $N = 1$ LLs. $m^*_{CF}$ is determined from the $B_\perp$ fits (color shaded line) based on the Pauli paramagnet model of a half-filled LL. Black dashed line shows theoretically predicted $P^*$ of the non-interacting CF with $m^*_{CF} = 0.6 m_e \sqrt{B_\perp}$ (*16*).

Given the dramatic mass enhancement of the CFs in the $N = 1$ LL, it is important to consider the meaning of $m^*_{CF}$. The conventional FQH states are built upon a "hard-core" model, where the short-distance interaction dominates over longer-range interactions (*1, 29*). The notion of $m^*_{CF}$ comes from short-range Coulomb energies generating gaps of $\hbar\omega^*_c = eB^*_\perp / m^*_{CF}$ in the low-energy spectrum (*2*). However, this conventional definition of $\hbar\omega^*_c$ and $m^*_{CF}$ breaks down when the short-range interactions are softened and are comparable to the longer-range interactions. In this case, conventional CF phases are unlikely to be stabilized (*30–32*) due to the stiffened residual interactions (see Fig. S12). Instead, the quasi-particles in the novel ground-states tend to be spin-polarized even at small $E_z$ (*33–35*). Such a scenario may be interpreted as the CFs having an enhanced $m^*_{CF}$ and a small $\hbar\omega^*_c$, as was assumed in prior experiments that interpreted data near $v = 5/2$ in terms of the cyclotron orbits of spin-polarized free CFs (*36, 37*). Even in this simplified interpretation, the gigantic value of $m^*_{CF}$ that we observed indicates markedly reduced short-range Coulomb energy and thus demonstrates the instability of the conventional CF phase. Therefore, the unexpectedly large $m_{CF}^*$ and the strikingly different $P^*$ phase diagram suggest substantially different correlation physics beyond the conventional model of the weakly interacting CFs in the $N = 1$ LL.

In summary, SRPT provides unprecedentedly detailed and high-fidelity determination of the ground-state spin-polarization of electronic systems, revealing surprises for the QH system. In the $N = 0$ LL, SRPT displays a spin phase diagram that mainly follows the non-interacting CF model but also provides evidence for the interacting CFs, such as the FQH skyrmion near $v = 1/3$ and the spin-unpolarized $v = 4/5$ and $6/5$ states. In the $N = 1$ LL, the results show full spin polarization that persists down to very low magnetic fields, deviating greatly from conventional CF model and demonstrating stronger longer-range correlations that can drive non-Abelian FQH states. Theory suggests that the QH system progresses from CF physics in the $N = 0$ LL (*2*), an intermediate and poorly understood regime in the $N = 1$ LL (*32*), to Hartree-Fock physics that produces stripe and bubble phases in the $N = 2$ and higher LLs (*38*). Future studies can reveal phase transitions in crystalline phases and also examine the validity of the Hartree-fock picture that predicts the non-uniform spin-polarization in higher ($N \geq 2$) LLs. Finally, we note that SRPT also lends itself to atomic layered systems by means of tunneling into exfoliated magnetic systems.

of a single quantum well: Spectroscopic evidence for domain formation. *Phys. Rev. B.* **70**, 075318 (2004).

**Acknowledgments:** We thank J. K. Jain and A. H. MacDonald for helpful discussions.; **Funding:** This work is supported by the Basic Energy Sciences Program of the Office of Science of the U.S. Department of Energy through contract no. FG02-08ER46514 and by the Gordon and Betty Moore Foundation through grant GBMF2931. The work at Princeton University was funded by the Gordon and Betty Moore Foundation through the EPiQS (Emergent Phenomena in Quantum Systems) initiative grant GBMF4420 and by the National Science Foundation MRSEC (Materials Research Science and Engineering Centers) grant DMR-1420541.; **Author contributions:** H.M.Y. and R.C.A. conceived the experiments. L. P., K.W.B, and K.W contributed in the epitaxial growth of the sample. H.M.Y. carried out the measurements. H.M.Y. and R.C.A. performed data analysis and prepared the manuscript. R.C.A. supervised the project.; **Competing interests:** Authors declare no competing interests.; **Data and materials availability:** The data that support the findings of this study are available from the corresponding authors on request.

**Supplementary Materials**

**Materials and Methods**

Our experiments were carried out using a floating-gate GaAs bilayer device, where the carrier density in each layer can be selectively tuned without a direct contact to the quantum wells (QWs). Detailed descriptions of the device can be found in Supplementary Section 1. The device consists of 180 Å and 280 Å thick GaAs QWs separated by a 130Å thick $Al_{0.25}Ga_{0.75}As$ tunnel barrier. The structure contains wide AlGaAs barriers above and below the QWs. The entire structure is sandwiched by heavily-doped $n^+$ GaAs electrodes.

To perform the pulsed tunneling measurements, we apply a pulsed voltage with 110 ∼ 400 ns of duration to inject electrons into the system under study (280 Å thick QW). We then measure the tunneling current by detecting the images charges of the tunneled electrons on the electrode. After the application of a short injection pulse, an opposite polarity of a discharge pulse and a minimum of 50 $\mu$s delay are applied to insure rethermalization of the system under study. All measurements were performed at a base temperature at $T$ = 30 mK.

**Supplementary Text**

1. Floating-gate memory device

In this work, we employed a floating-gate memory device (*39*). The floating-gate structure has two advantages. First, the floating-gate structure does not require a direct contact to the QWs and therefore allows simple device fabrication. In order to tune the total carrier density, we apply a large gate voltage to trap or to leak the charges residing in the double QWs. We then apply a smaller voltage at the same gate to partition the charges between the two QWs. Second, the heavily-doped electrodes on the top and bottom surface eliminate the long-range Coulomb disorders. Previous work (*40*, *41*) demonstrated that the stability of the even-denominator FQH states weakly depends on the transport mobility. Instead, the long-range fluctuations arising from the unscreened ionized impurities in the doped layers are the dominant factor limiting the energy gap of the non-abelian ground states. In our floating-gate structure, any electric charge perturbations from the doped layers are screened within few angstroms in the AlGaAs barrier (see Fig. S1). The long-range correlations in the floating-gate device are thus expected to be more stable against the Coulomb disorders in the doped layers.

In the bilayer device, there are several possible effects of placing the second 2DES nearby the system under study. A previous experiment (*42*) showed that the second 2DES screens the electron-electron interactions and softens the FQH energy gap. However, in the SRPT measurements, the second 2DES is always fixed at the incompressible $\nu$ = 1 state. Thus, such screening effect would likely be greatly suppressed because the immobile charges in the $\nu$ = 1 insulator cannot screen the interactions in the system under study. Interlayer excitonic interactions (*43*) can also modify the intra-layer interactions in the system under study. Although the excitonic interactions cannot completely be ruled out at extremely low magnetic fields, such effects are very weak when, as in our case, the two layers are widely separated and the sum of the LL filling factors of the two 2DESs differs from one.

## 2. Detailed comparison between SRPT and other techniques

Despite the progress of NMR, optical, and transport measurements (*3–7*), there have been substantial experimental challenges that hampered investigation of the spin-dependent phenomena in the QHE regime. In optical techniques, photoexcited holes favor the formation of skyrmions that depolarize the ground-state spin in the $N = 1$ LL (*44–46*). Transport methods cannot measure the spin-polarization of skyrmion ground-state nor the partially filled $\Lambda$-levels, where our work now demonstrates significant deviations from the non-interacting CF model. Also, some transport measurements rely on tilting the applied field, and this may drive spin-independent phase transitions that confound the results (*47*). Resistively-detected NMR spectroscopy (RD-NMR) (*28*) involves RF-heating and substantial changes to the carrier density that can alter the measured spin state (*48*). The nonuniform response also makes the analysis of RD-NMR spectra difficult (*49*). Finally, the weak hyperfine interactions limits the applications of RD-NMR to other materials, such as graphene and *p*-type GaAs.

SRPT provides an alternative direct approach to probe the spin-properties of a QH system. The short duty cycles of the pulsed tunneling measurement prevent the undesirable perturbations such as RF-heating or photoexcited defects that commonly occur with prior methods. A tunneling pulse width is also set much shorter than the RC charging time to minimize other possible effects that can alter the ground-state (see Fig. S2A). In addition, the simple integration of the *I-V* curve gives a quantity directly proportional to the spin-polarized carrier density, whereas the deduced value of the spin-polarization in RD-NMR depends on a fit that demands a knowledge of the bound-state wavefunction and the spatial electronic structure (*28*).

In SRPT, the use of the $\nu = 1$ ferromagnet (the strongest ferromagnetic state in a QH system) allows the precise measurements at low magnetic fields. The lower $B_\perp$ limit is set by the polarization of the $\nu = 1$ ferromagnet. At a very small $B_\perp$, the strength of the Coulomb exchange interactions is significantly reduced, and thermally excited spin-waves weaken the polarization at $\nu = 1$. Figure S3B shows the ratio ($\gamma$) of depolarized electrons in the $\nu = 1$ QH ferromagnet as a function of $B_\perp$. We deduce $\gamma$ from the integrated tunnel current when both the 2DES layers are at $\nu = 1$. The $B_\perp$ dependence of $\gamma$ reveals an Arrhenius behavior, where a decrease in the exchange Coulomb energy leads to an exponential growth of the spin-flips. Below $B_\perp = 1$ T, $\gamma$ rapidly increases and limits the SRPT measurement.

## 3. 2D - QH ferromagnet tunneling model

Tunneling current between two parallel layers are expressed as follows

$$I(V) = \frac{4\pi e}{\hbar} \sum_{k,k'} \int dE \int dE' \left| M_{k,k'} \right|^2 D(E) D'(E') \delta_{E+eV,E'}$$
$$\times \left[ f(E,T)(1 - f'(E',T)) - f'(E',T)(1 - f(E,T)) \right],$$

(S1)

where $D(E)$ and $D'(E)$ are the density of states of each 2DES.

The Fermi functions for each layer are given by $f(E,T) = \left\{ 1 + e^{|E-\mu|/k_B T} \right\}^{-1}$ and

$f'(E,T) = \left\{ 1 + e^{|E-\mu'|/k_B T} \right\}^{-1}$. The tunneling matrix element $M_{k,k'}$ is defined by

$$M_{k,k'} = -\frac{\hbar^2}{2m_e^*} \left\{ \psi(z)^* \frac{\partial \psi'(z)}{\partial z} - \psi'(z)^* \frac{\partial \psi(z)}{\partial z} \right\} \delta_{k_s,k'_s} \int \varphi_{N,X}(y) \varphi_{N',X'}(y) dy, \quad \text{(S2)}$$

where $\psi(z)$ is the out-of-plane wavefunction perpendicular to the 2DES and $\varphi_{N,X}(y)$ is the in-plane $N^{th}$ LL wavefunction centered at position $x = X$.

The first term in Eq. S2 is the matrix element perpendicular to the 2DES. At a fixed electron density, the perpendicular component is nearly constant over the tunneling energy range of the measurements (9). The last integral term is the in-plane component of the matrix element. The in-plane component is governed by the overlap integral between the two LL wavefunctions and stays constant at a fixed $B_{\parallel}$. Detailed description of the wavefunction overlap integral can be found in Supplementary Section 5.

In the SRPT measurements, one layer is always fixed at the ferromagnetic $\nu = 1$ state. For the 2D-QH ferromagnet tunnel junction, only the spin-down electrons are allowed to tunnel into the $\nu = 1$ QH ferromagnet. Also, the following condition holds at low temperatures: $f(E,T)(1-f'(E',T))\delta_{E+eV,E'} \gg f'(E',T)(1-f(E,T))\delta_{E+eV,E'}$. Thus, the tunneling $I$-$V$ characteristics of 2D-QH ferromagnet is given by the following expression:

$$I(V) = \frac{4\pi e}{\hbar} \sum_{k,k'} \left| M_{k,k'} \right|^2 \int dE \int dE' D_\downarrow(E) D'_\downarrow(E') \delta_{E+eV,E'} \left[ f(E,T)(1-f'(E',T)) \right], \quad \text{(S3)}$$

where $D(E)$ and $D'(E)$ are the spin-down density of states of the system under study and the $\nu = 1$ QH ferromagnet, respectively. Integrating the tunneling current in Eq. S3 yields

$$\int I(V) dV = \frac{4\pi e}{\hbar} \sum_{k,k'} \left| M_{k,k'} \right|^2 \int dE D_\downarrow(E) f(E,T) \int dV D'_\downarrow(E+eV)(1-f'(E+eV,T)) \quad \text{(S4)}$$

The last two integral terms are equal to the occupied spin-down density in the system under study ($n_\downarrow$) and the unoccupied spin-down density in the $\nu = 1$ QH ferromagnet, respectively. The unoccupied spin-down density at $\nu = 1$ always equals to LL degeneracy $n_D$.

The integrated tunneling current in Eq. S4 can be simplified as follows:

$$\int I(V) dV = \frac{4\pi e}{\hbar} \sum_{k,k'} \left| M_{k,k'} \right|^2 n_\downarrow n_D \quad \text{(S5)}$$

where $\sum_{k,k'} \left| M_{k,k'} \right|^2$ is the tunneling matrix element. It follows from Eq. S5 that the integrated tunneling current in 2D-QH tunnel junction gives a quantity proportional to the occupied spin-down density ($n_\downarrow$) in the system under study.

## 4. Tunneling matrix element calibration

We calibrate the SRPT spectra by dividing the measured current by our estimate of $\sum_{k,k'}|M_{k,k'}|^2$. In a 2DES with finite layer-thickness, the bound-state wavefunction shifts in response to the applied DC electric field. Thus, the out-of-plane component of $M_{k,k'}$ is a slowly varying function of $v$ at a fixed $B_\perp$.

There are two possible ways to deduce the matrix element. First, as shown by Eq. S3, the matrix element is independent of the spin-polarization. By measuring the spin-independent but $v$-dependent tunneling current, we can deduce $M_{k,k'}$. To perform spin-degenerate tunneling, we first change the $v=1$ ferromagnet to a trivial $v=0$ state and then measure the $v$ dependence of the current flowing into the spin-degenerate LL (Fig. 1B). In this way, we can measure the contribution of $M_{k,k'}$ as a function of $v$. This method, however, demands separate measurements at different DC gate voltages. In the second method, we determine $M_{k,k'}$ from the tunneling $I$-$V$ curve measured at the opposite tunneling energy, but at the same DC gate voltage. While the integrated tunneling current below $E_F$ is proportional to $\sum_{k,k'}|M_{k,k'}|^2 n_\downarrow$, the integrated tunneling current above $E_F$ is proportional to the spin-up hole density multiplied by the matrix element $\sum_{k,k'}|M_{k,k'}|^2(n_D-n_\uparrow)$. The difference of the integrated current measured above and below $E_F$ equals to $\sum_{k,k'}|M_{k,k'}|^2(n_D-n_\uparrow-n_\downarrow)=\sum_{k,k'}|M_{k,k'}|^2(n_D-v)$. For each $v$, we can therefore determine $\sum_{k,k'}|M_{k,k'}|^2$ using the simple algebraic expression.

Figure S5B shows the bare polarization curve determined from the raw current (black) and the calibrated current (red). We used the second method described above to calibrate the tunneling current in this paper. At large $v$, the uncorrected polarization curve exceeds the maximally polarized curve (gray). As explained earlier, the matrix element decreases as $v$ increases. The decrease in the matrix element reduces the spin-down tunneling current and therefore increases the spin-polarization more than the maximum polarization. Our calibration method compensates for this spin-independent matrix element effect. The polarization curve deduced from the calibrated current does not exceed the maximal polarization curve. We also note that the application of matrix element correction does not change the qualitative features such as the peaks and dips in the polarization curve.

## 5. The effect of the Landau index ($N$) selection rule in 2D-2D tunneling

Tunneling allows measurements of the electronic density of states (DOS). In a 3D-2D tunnel junction employed in our previous work (9), the differential conductance $dI/dV$ spectra measures the DOS of a 2DES in a magnetic field. Spectra taken in the presence of $B_\perp$ show multiple staircase patterns of the equidistant LLs as a function of $v$ (Fig. S6B). In moving to a 2D-2D tunnel junction, the selection rule imposed by in-plane momentum conservation substantially modifies the tunneling spectrum. In contrast to 3D-2D spectra, the 2D-2D tunneling spectra in Fig. S6C shows a single staircase pattern. This single staircase pattern indicates that there is only one energy that allows tunneling for any given $v$. This condition occurs when the LLs with the same orbital index $N$ are aligned at the same energy. Otherwise, tunneling into the LLs with different $N$ is forbidden.

The $N$ selection rule in 2D-2D tunneling can be understood as follows. In the semiclassical picture, free electrons residing in the QWs move on cyclotron orbits under $B_\perp$. These semiclassical orbits can be understood as quantized harmonic oscillator in which electrons are confined in parabolic magnetic potentials (see Fig. S7A). The transition probability between the LLs is proportional to the square of the spatial overlap between the initial and final state wavefunctions. In the absence of $B_\parallel$, a transition that does not conserve $N$ is forbidden because the wavefunctions are mutually orthogonal. The selection rule of conserving $N$ is, however, broken in a tilted magnetic field due to a "momentum boost" $\hbar\Delta k_y = eB_\parallel d$ obtained by tunneling electrons, where $d$ is the physical distance between the two QWs. The additional momentum displaces the centers of the cyclotron orbits in the two QWs relative to each other by $\Delta X = B_\parallel d / B_\perp$. Non-zero overlap integrals between the two displaced harmonic oscillator wavefunctions allow tunneling with non-conservation of $N$. Therefore, an idealized 2D-2D tunneling device shows a single peak in its $I$-$V$ characteristic at $B_\parallel = 0$. Multiple current peaks that correspond to individual inter-LL tunneling appear at $B_\parallel > 0$ (Fig. S7B).

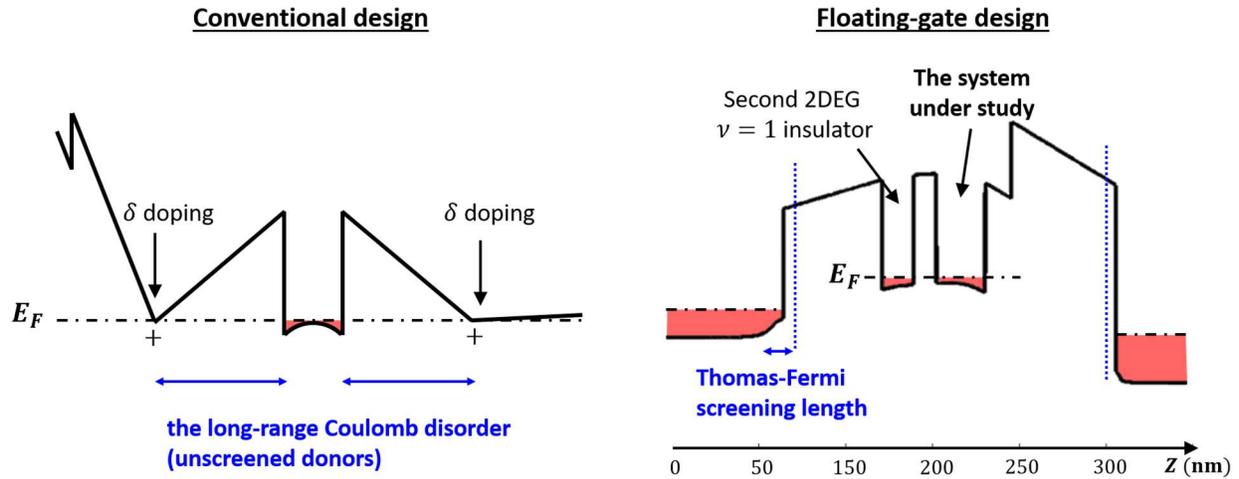

**Fig. S1. Comparison between the conventional modulation doped (left) and the floating-gate memory (right) structure.** The conventional structure requires the Si doping layers that induce the long-range fluctuations in the QW. In the floating-gate structure, the charge fluctuations in the doped layer exponentially decay over the Thomas-Fermi screening length and cannot influence the system under study.

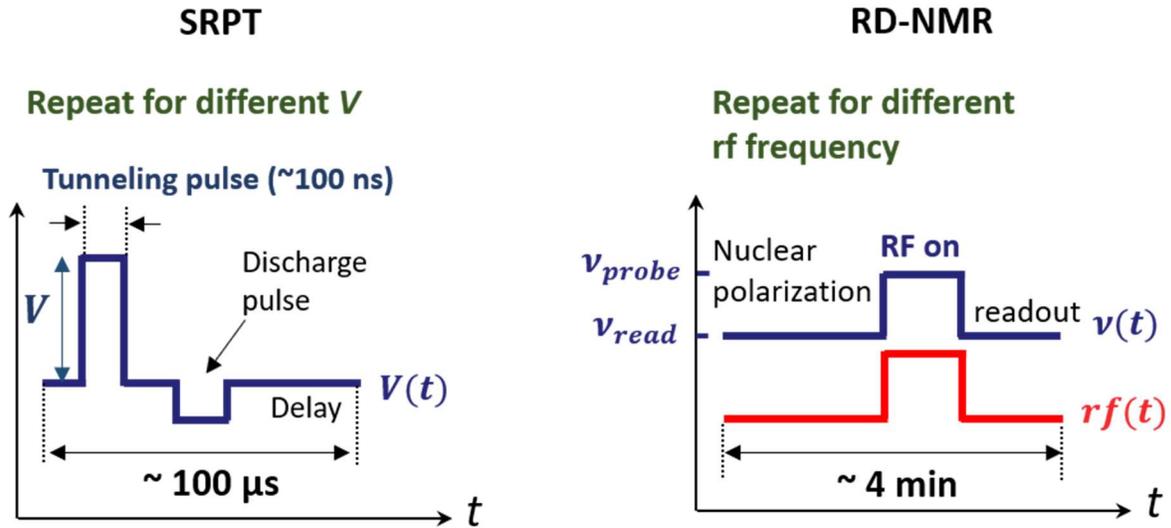

**Fig. S2. Comparison between SRPT (left) and RD-NMR (right) measurement sequence.**
SRPT uses a short pulse sequence that consists of a 100 ns tunneling pulse, an opposite discharge pulse, and a 50 µs delay. RD-NMR measurement employs a longer process of nuclear polarization, rf radiation, and readout (*28*). In SRPT, spin-polarized tunnel current (*I*) is measured as a function of tunneling pulse height (*V*). In the RD-NMR, the change of longitudinal resistance ($\Delta R_{xx}$) is measured as a function of RF frequency.

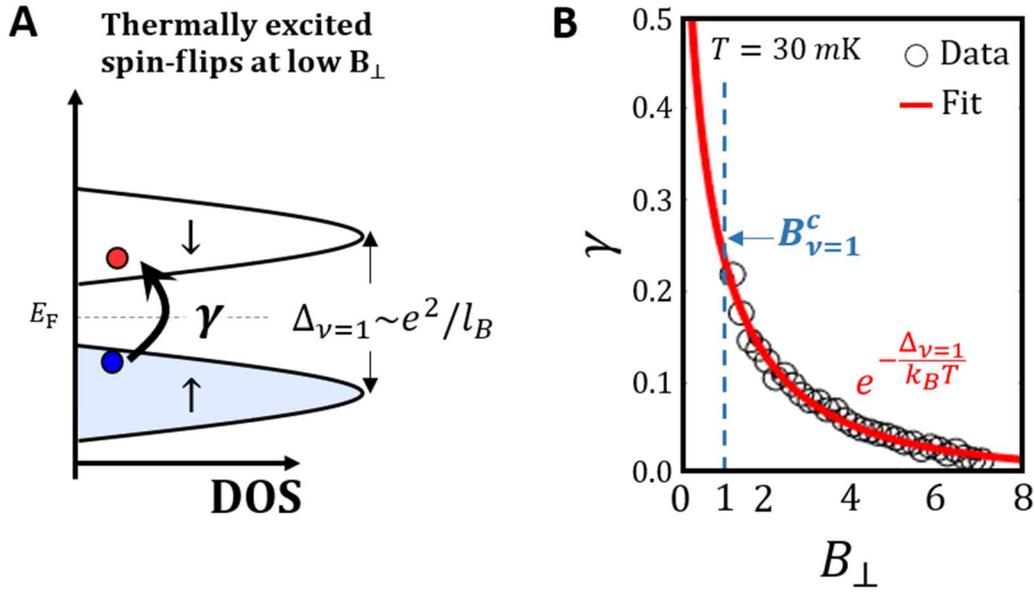

**Fig. S3. The lower $B_\perp$ limit of SRPT measurements. (A)** A cartoon illustrating spin-flips in the $\nu = 1$ QH ferromagnet at small $B_\perp$. The spin energy gap $\Delta_{\nu=1}$ is proportional to the exchange interaction that grows as $e^2/l_B$ and $\sqrt{B_\perp}$. When the spin gap decreases at low $B_\perp$, spin waves are thermally excited and reduce the spin-polarization. **(B)** The ratio ($\gamma$) of depolarized electrons in the QH ferromagnet as a function of $B_\perp$ at a fixed $T = 30$ mK. At $B_\perp$ below 1 T, $\gamma$ rapidly increases and limits the SRPT measurements.

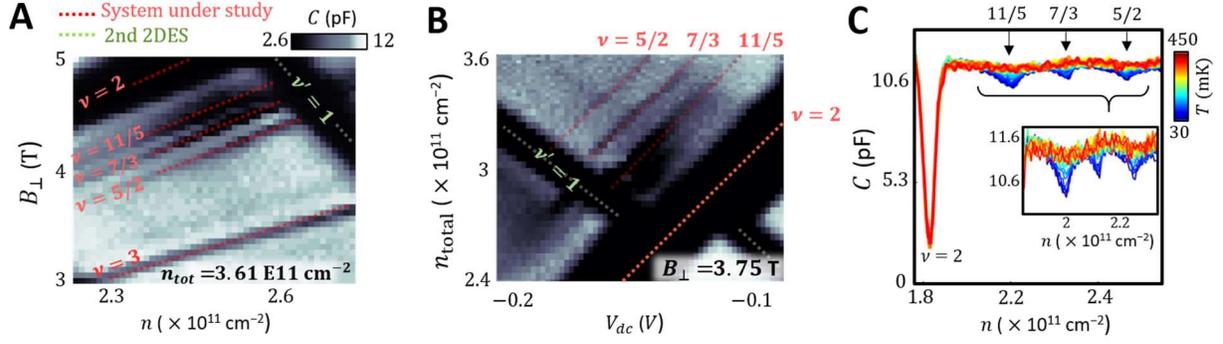

**Fig. S4. Low frequency magnetocapacitance in the $N = 1$ LL. (A)** $B_\perp$ dependence of sample capacitance measured at a fixed total density $n_{\text{tot}} = 3.61$ E11 cm$^{-2}$. A low frequency ac excitation ($f = 17$ Hz and $V_{ac} = 265$ μV) is applied. Dashed lines are guides to the eye showing the filling factors of the system under study (red) and the second 2DES (green). **(B)** $n_{\text{tot}}$ dependence of low frequency ($f = 28$ Hz) capacitance measured at a fixed $B_\perp = 3.75$ T. Horizontal axis is the DC gate voltage $V_{dc}$ that partitions the total carriers ($n_{\text{tot}}$) into the two layers. **(C)** Temperature dependence at a fixed $B_\perp = 3.75$ T and a fixed total density $n_{\text{tot}} = 3.38$ E11 cm$^{-2}$.

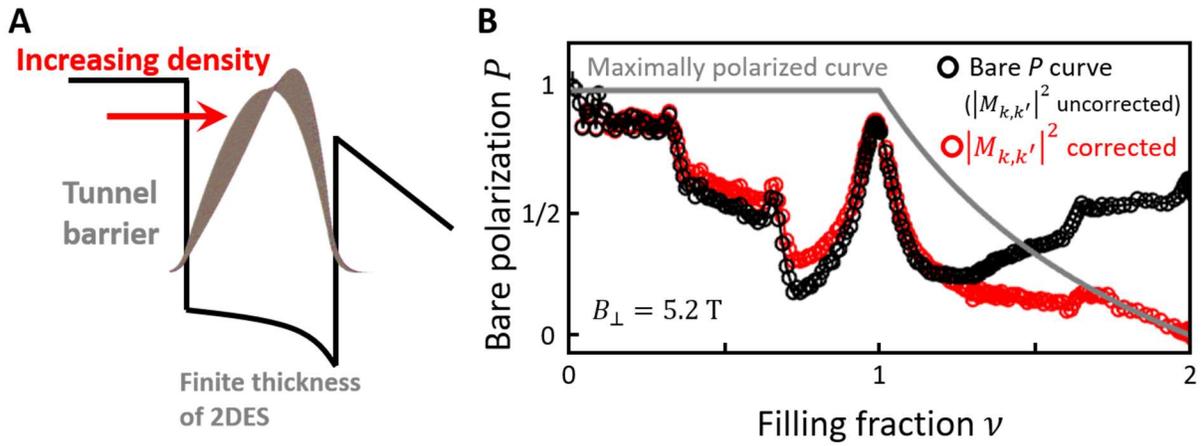

**Fig. S5. Tunneling matrix element calibration. (A)** Due to the finite thickness of the QW, the wavefunction moves away from a tunnel barrier as the electron density increases. The tunneling matrix element gradually changes as $\nu$ increases at a fixed $B_\perp$. **(B)** Comparison between uncorrected (black) and corrected (red) $P$ curves. The matrix element correction compensates for the out-of-plane wavefunction shift in a QW. Note that the application of matrix element correction does not change the qualitative features such as the peaks (maxima) and dips (minima).

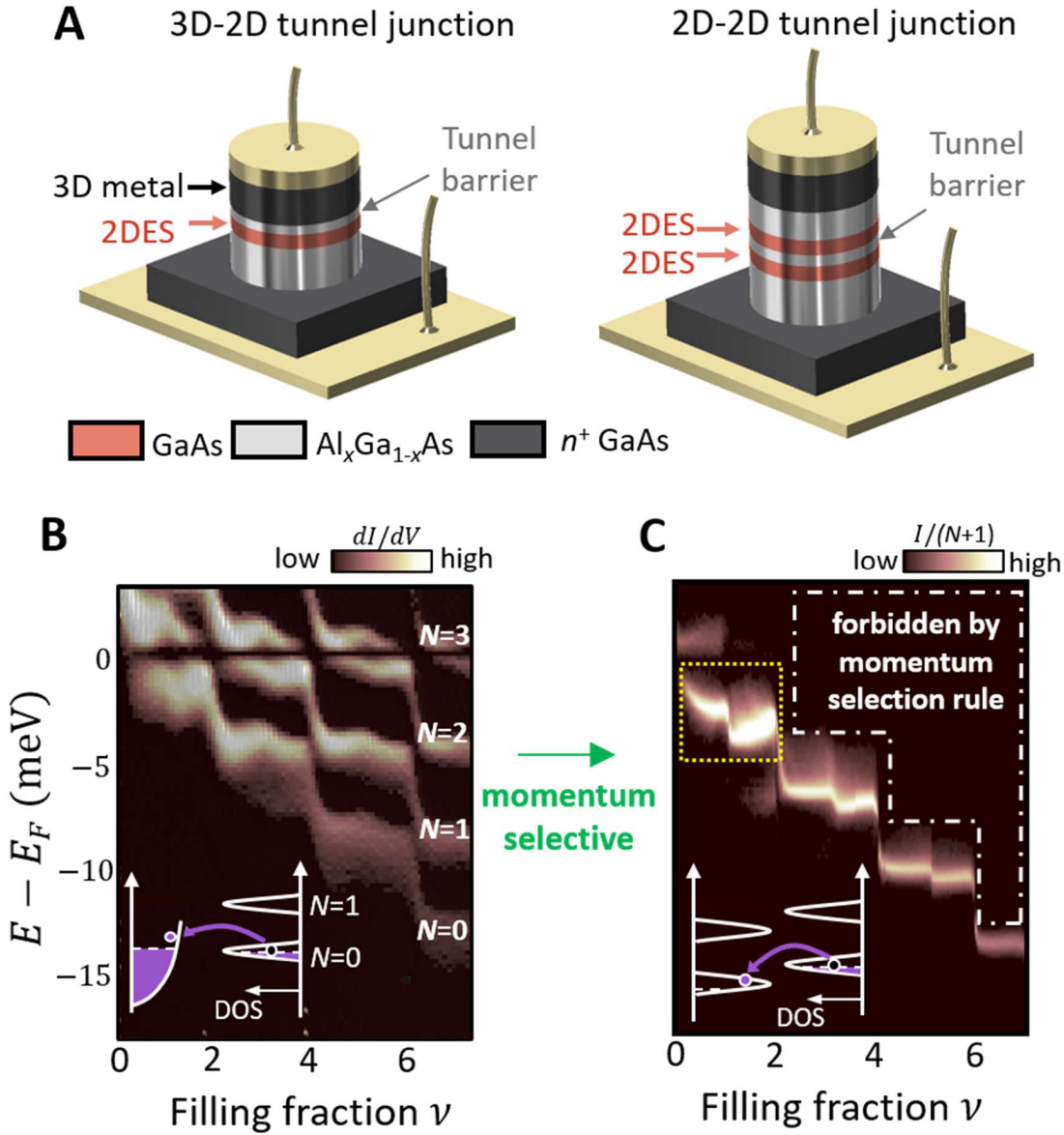

**Fig. S6. Comparison between 3D-2D and 2D-2D tunnel junctions. (A)** Schematic views of 3D-2D and 2D-2D tunnel devices consisting of GaAs QWs and a AlGaAs tunnel barrier. **(B and C)** Tunneling spectra measured at a fixed $B_\perp = 2$ T. Insets show the schematic energy diagram for the (B) 3D-2D and (C) 2D-2D tunnel junctions. Vertical axis is the energy $E$ referenced to the Fermi energy $E_F$. The horizontal axis is $\nu$. The color scale in the 3D-2D spectra is proportional to $dI/dV$, whereas the color scale in the 2D-2D spectra is $I$ scaled by the topmost occupied LL index ($N + 1$). **(B)** In the 3D-2D tunnel junction (*9*), the $dI/dV$ measures the DOS of the 2DES under study. The $dI/dV$ spectra show multiple staircase patterns of LLs as a function of $\nu$. **(C)** In the 2D-2D tunnel junction, quantized energy levels develop in both 2D systems. The 2D-2D tunneling spectra show a single staircase pattern due to the LL index selection rule (see Supplementary Section 5 for a detailed explanation).

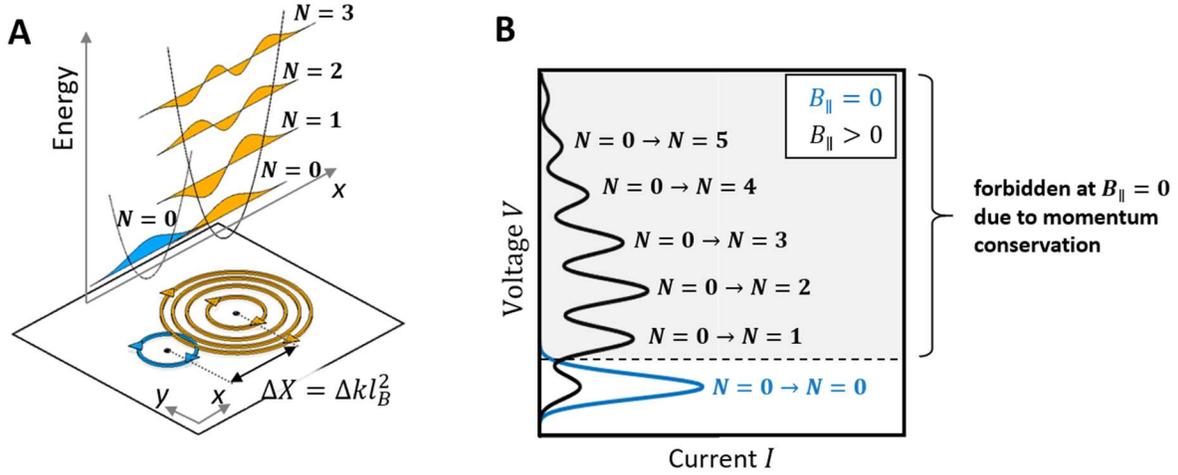

**Fig. S7. The Landau index _N_ selection rule for 2D-2D tunneling. (A)** A simple cartoon model describing the effect of _N_ selection rule. In the presence of $B_\perp$, 2D electrons move on cyclotron orbits. These semiclassical orbits can be understood as a quantized harmonic oscillator. An additional in-plane field $B_\parallel$ displaces the centers of the cyclotron orbits as well as the harmonic oscillators by $\Delta X = B_\parallel d / B_\perp$. The inter-LL transitions depend on the overlap integral of the two displaced harmonic oscillator wavefunctions. **(B)** Calculated tunneling _I-V_ characteristics in both $B_\perp$ and $B_\parallel$. At $B_\parallel = 0$ (blue), there is a single tunneling peak that corresponds to _N_-conserving tunneling. Inter-LL tunneling that does not conserve _N_ is forbidden because the wavefunctions are orthogonal. On the other hand, tunneling with non-conservation of _N_ is allowed at $B_\parallel > 0$ as a result of the non-zero overlap integral between the two displaced wavefunctions (black).

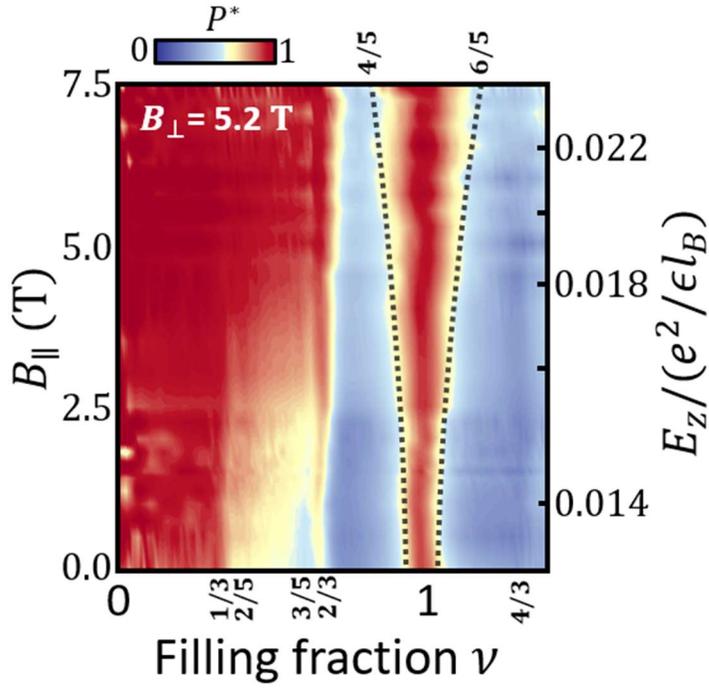

**Fig. S8. The $P^*$ phase diagram measured as functions of $B_∥$ and $\nu$ at a fixed $B_⊥$ = 5.2 T.** Color scale is adjusted to highlight the contrast. Dashed lines are guides to the eye. Note that the IQH skyrmion is suppressed at the largest applied $B_∥$, whereas the FQH skyrmion near $\nu$ = 1/3 is already suppressed at $B_∥$ = 0 T.

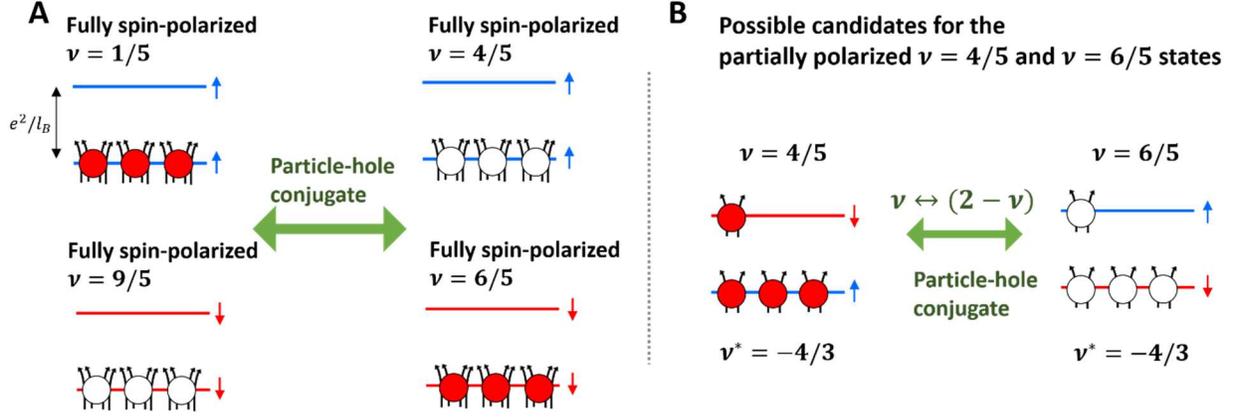

**Fig. S9. Possibility of the interacting CF phases at $2/3 \leq \nu \leq 4/5$ and $6/5 \leq \nu \leq 5/3$. (A)** Fully polarized non-interacting CFs at $\nu = 1/5$, $4/5$, $4/5$, and $9/5$. The $\nu = 1/5$ FQH states correspond to the $\nu^* = 1$ IQH state of the four-flux CFs. When the LLs are spin-polarized at large $B$, the particle-hole conjugate FQH state at $\nu = 4/5$ is expected in the spin-up branch of the $N = 0$ LL. Similarly, the four-flux FQH state at $\nu = 9/5$ and its particle-hole conjugate at $\nu = 6/5$ are expected in the spin-down branch. Thus, the FQH states at $\nu = 4/5$ and $\nu = 6/5$ are expected to be fully spin-polarized in this model. **(B)** A possible candidate for the observed partially spin-polarized $\nu = 4/5$ and $\nu = 6/5$ FQH states. As shown in Fig. 3A, the measured $P^*$ near $\nu = 4/5$ and $\nu = 6/5$ are not fully polarized. This behavior cannot be understood in terms of the non-interacting model. One possible alternative model (*14, 15*) is interacting two-flux CFs (not four-flux CFs) at $2/3 < \nu \leq 4/5$ and $6/5 \leq \nu < 5/3$. For instance, the $\nu = 4/5$ and $\nu = 6/5$ states correspond to the $\nu^* = -4/3$ states of the two-flux CFs (the negative sign of $\nu^*$ represents the negative effective magnetic field). When the CFs form the $\nu^* = -4/3$ FQH state, the partially spin-depolarized FQH states are allowed at $\nu = 4/5$ and $\nu = 6/5$. The calculated $P^*$ phase diagram in Fig. 3B assumes these interacting two-flux CFs with the same $m_{CF}^*$ at $2/3 < \nu \leq 4/5$ and $6/5 \leq \nu < 5/3$.

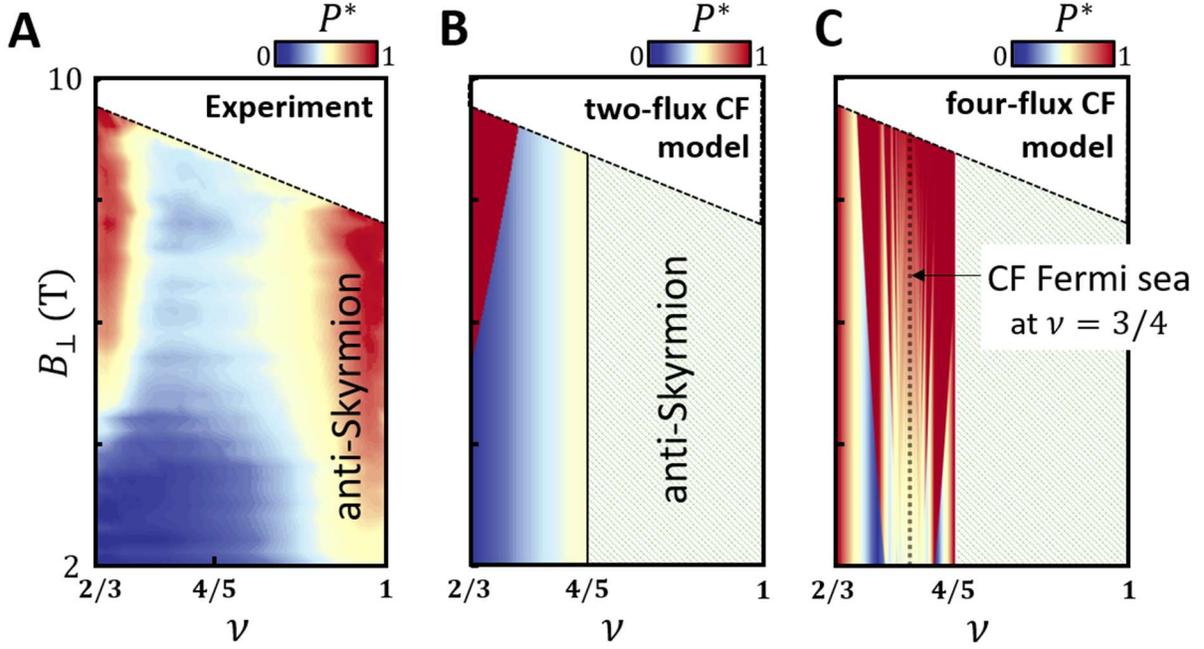

**Fig. S10. Comparison of measured and calculated spin-polarization at $2/3 < \nu \leq 4/5$. (A)** The selected region of the $P^*$ phase diagram in Fig. 3A. **(B and C)** Spin-polarization of the two-flux CF phases is calculated using an effective mass $m^*_{CF} = 0.66 m_e \sqrt{B_\perp}$. For the four-flux CF phases, an effective mass $m^*_{CF} = 0.34 m_e \sqrt{B_\perp}$ deduced from a prior transport measurement (*50*) is assumed. The calculated $P^*$ diagrams in (B) and (C) show a contrasting behavior of the two-flux and four-flux CFs. The four-flux CF state at $\nu = 4/5$ is fully polarized, whereas the two-flux CF state at $\nu = 4/5$ is spin-unpolarized even at the largest applied $B_\perp$. The measured $P^*$ (as well as the $B_\parallel$ dependence of $P^*$ in Fig. S8) shows a similar trend to the two-flux CF model. The $\nu = 4/5$ state and the nearby states are spin-unpolarized over a broad range of $B_\perp$ and $B_\parallel$. The agreement between the measured $P^*$ and the two-flux CF model suggests that the spin-unpolarized FQH states at $2/3 < \nu \leq 4/5$ and $6/5 \leq \nu \leq 4/3$ are likely to be the FQH states of interacting two-flux CFs (Fig. S9B).

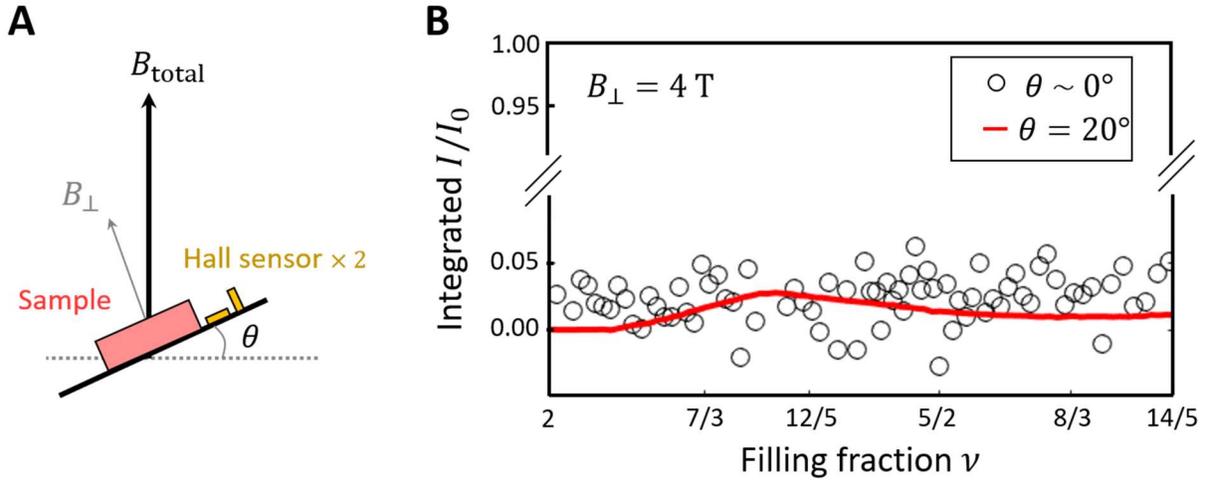

**Fig. S11. (A)** Schematic showing sample tilting. Two perpendicular Hall sensors are used to measure the in-plane and out-of-plane components of an applied magnetic field. **(B)** Integrated tunneling current in the $N = 1$ LL measured at different tilt angles $\theta \sim 0°$ and $\theta = 20°$. Due to the $N$ selection rule, the tunneling current and the signal-to-noise ratio are substantially reduced at $\theta = 0°$. Although determination of $P^*$ becomes challenging near $\theta = 0°$, the small integrated current near $\nu = 5/2$ and $\nu = 8/3$ indicates full spin-polarization and is consistent with the result at $\theta = 20°$. Theory also predicts that the effect of $B_\parallel$ on the electron interactions is not significant at a small $\theta$ (*51*).

**Conventional CF phases**          **Breakdown of CF phases**

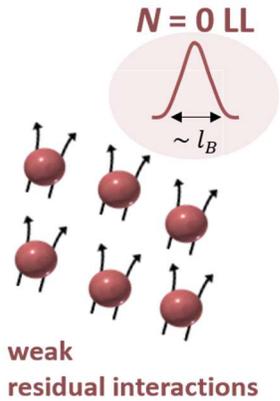 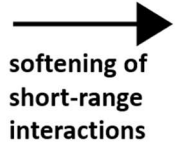 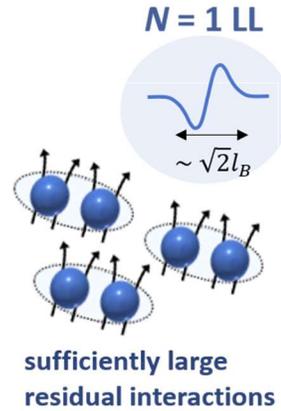

**Fig. S12. A cartoon comparing the wavefunctions (inset) and interparticle correlations in the $N = 0$ and $N = 1$ LLs.** In the $N = 0$ LL, the strong short-range repulsive interactions favor the conventional FQH states (see Fig. 2B). In contrast, the larger spatial extent of the LL wavefunction in the $N = 1$ LL causes softening of the short-distance part of the interactions. When the strength of the short-range interactions is comparable to the longer-range interaction energies, the conventional FQH state becomes energetically unstable due to the residual inter-CF interactions (*52*).